\documentclass[aps,prl,twocolumn,groupedaddress,showpacs]{revtex4-1}
\usepackage{amsmath,amssymb,bm,color,graphicx,ulem}

\begin{document}

\title{
Triangular Antiferromagnet with Nonmagnetic Impurities}
\author{V. S. Maryasin and M. E. Zhitomirsky}
\affiliation{
Service de Physique Statistique, Magn\'etisme et Supraconductivit\'e,
UMR-E9001 CEA-INAC/UJF, 17 rue des Martyrs, 38054 Grenoble, France
}

\begin{abstract}
The effect of nonmagnetic impurities on the phase diagram of the classical
Heisenberg antiferromagnet on a triangular lattice is investigated.
We present analytical arguments confirmed by numerical calculations that
at zero temperature vacancies stabilize a conical state providing an example
of ``order by quenched disorder'' effect. Competition between thermal
fluctuations and the site disorder leads to a complicated $H$--$T$ phase
diagram, which is deduced from the classical Monte Carlo simulations for a
representative vacancy concentration. For the  $XY$  triangular-lattice
antiferromagnet with in-plane external field nonmagnetic impurities stabilize
the fan-like spin structure. We also briefly discuss the effect of quantum
fluctuations.
\end{abstract}


\date{19 November, 2013}

\maketitle

{\it Introduction.}---%
Spin vacancies produced by substitution of nonmagnetic ions is a common form
of disorder in magnetic solids. Nonmagnetic impurities are often used experimentally
as a probe of local spin correlations \cite{Vajk02,Bobroff09}. Accordingly, many
theoretical works were devoted to investigation of magnetization patterns and spin
textures around a single impurity in ordered, spin-liquid, and quantum-critical
antiferromagnets
\cite{Sigrist96,Sachdev99,Sachdev03,Sushkov03,Hoglund07,Eggert07,Chen11,Sen11,Wollny11,%
Weber12,Wollny12,Brenig13}.
Scaling from a single vacancy to a more realistic situation of small but finite
concentration of impurities is straightforward for simple collinear antiferromagnets.
In the case of frustrated magnets with a ground-state degeneracy the problem of
collective impurity behavior becomes much more nontrivial due to a possible
``order by quenched disorder'' effect \cite{Villain79,Henley89,Savary11}.

\begin{figure}[b]
\includegraphics[width=0.9\columnwidth, keepaspectratio]{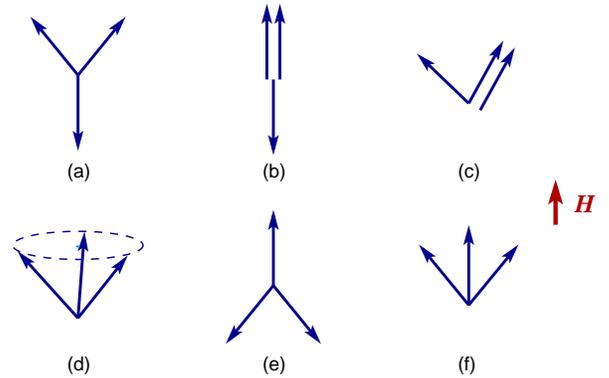} 
\caption{Ordered magnetic states of a TAFM in an external field. Spin configurations appearing
for the TAFM without impurities: (a) coplanar $Y$-state, (b) collinear $uud$ state,
and (c) coplanar 2:1 ($V$) state. Spin configurations in the presence of nonmagnetic
impurities: (d) conical (umbrella) state of the Heisenberg TAFM,
(e) anti-$Y$ state and equivalent (f) fan state of the $XY$ TAFM.
}
\label{fig:conf}
\end{figure}

In this Letter we consider the triangular-lattice antiferromagnets (TAFMs), which attracted
a lot of interest in the past as a paradigmatic example of geometrical frustration
\cite{Lee84,Kawamura85,Chubukov91,Balents10,Struck11} and also due to their intrinsic
multiferroicity \cite{Katsufuji01,Kenzelmann07}. A single nonmagnetic impurity embedded into
the TAFM was investigated by Wollny {\it et al.}\ \cite{Wollny11}. They found that a fractional
magnetic moment collinear with the ``missing'' spin is formed around a vacancy site.
In magnetic field the clean classical TAFM exhibits an ``accidental'' degeneracy consisting
in an arbitrary orientation of the spin plane with respect to the field direction
\cite{Kawamura85,Chubukov91}. Hence, the impurity moment may stabilize the same coplanar
magnetic structures, Figs.~\ref{fig:conf}(a)--(c), that are also favored by thermal and quantum
fluctuations.

Below, we demonstrate analytically and numerically that the single-impurity scenario breaks
down in the case of the TAFM already at very small vacancy concentrations. We find that in
magnetic field impurities favor the {\it least} collinear state,
{\it i.e.}, the non-coplanar conical or umbrella spin structure, Fig.~\ref{fig:conf}(d).
The behavior of the diluted TAFM is, therefore,
strongly affected by competition between quenched and thermal disorder.
The phase diagram of a diluted classical TAFM provides a rare  physical example in which
nonmagnetic impurities tune bulk properties of an ordered antiferromagnet and
drastically modify its phase diagram.

{\it Theory.}---%
We consider the classical Heisenberg antiferromagnet on a depleted
triangular lattice described by the spin Hamiltonian
\begin{equation}
\hat{\cal H} = J \sum_{\langle ij\rangle} p_i p_j\,\mathbf{S}_{i}\cdot
\mathbf{S}_{j} - {\bf H}\cdot \sum_{i}p_i{\bf S}_i
\label{H}
\end{equation}
with $|{\bf S}_i|=1$ and $p_i = 1$ or 0 for filled and empty sites, respectively.
The clean model with no depletion, $p_i\equiv 1$, orders at $T=0$ in
the $120^\circ$ three-sublattice magnetic structure described by the
wave vector $\mathbf{Q}=(4\pi/3,0)$. In magnetic field the classical energy
is minimized for spin configurations constrained by the magnetization
of each triangular plaquette:
\begin{equation}
\mathbf{S_{\bigtriangleup}} = \mathbf{H}/(3J) \ .
\label{Triangle}
\end{equation}
The constraint leaves undetermined the orientation of the spin plane
and sublattice directions inside that plane. This degeneracy
persists up to the saturation field $H_{s}=9J$.

An empty site produces a strong local perturbation in the $120^\circ$ magnetic
structure leading to readjustment of neighboring spins with a characteristic
power-law decay with distance \cite{Wollny11}. Therefore, for the sake of
analytical analysis we resort to a simpler model of weak bond disorder:
the lattice is assumed to be fully filled, whereas the exchange parameters
$J_{ij}$ fluctuate randomly about the mean value $J$ with
$\langle \delta J_{ij}^2\rangle = \delta J^2$.
The bond disorder may develop in magnets with sizeable magnetoelastic coupling.
In our case, a transformation between the site-disorder model (\ref{H})
and the bond-disorder model with $J_{ij}=Jp_ip_j$ can be constructed  by
(i) allowing $p_i\in (0,1)$, {\it i.e.}, letting spins on impurity sites change their
length continuously, and (ii) assuming sufficient density
of impurities such that coupling constants for adjacent bonds fluctuate independently.
The latter condition is expected to be satisfied for $n_{\rm imp}\agt 3$--5\%
once a distance between impurities is not too large. Numerically, we find
that the qualitative behavior obtained for the bond-disorder model remains
valid for the site-dilution model (\ref{H}) with the vacancy density as low as
$n_{\rm imp}\sim 0.1$\%.

To treat the effect of small thermal fluctuations $T\ll J$ and weak quenched
disorder $\delta J\ll J$ on equal footing, we employ the real-space perturbation
theory used  for clean frustrated magnets in \cite{Long89,Heinila93,Canals04}.
The starting point is an arbitrary ground-state spin configuration
of the TAFM in external magnetic field. To take into account small fluctuations
the Heisenberg Hamiltonian is transformed into the ``rotating'' local frame
with the $z_i$ axis directed parallel to ${\bf S}_i$:
\begin{eqnarray}
\hat{\cal H} & = & \sum_{\langle ij\rangle} J_{ij} \bigl[  S_i^y S_j^y + \cos\theta_{ij}
\bigl(S_i^z S_j^z + S_i^x S_j^x \bigr)
\nonumber \\
&& \mbox{\quad} + \sin\theta_{ij} \bigl(S_i^z S_j^x - S_i^x S_j^z \bigr) \bigr]
   - \mathbf{H}\cdot\sum_i\mathbf{S}_i \ .
\label{Hloc}
\end{eqnarray}
Here and below $S_i^\alpha$ denote spin components in the local frame and
$\theta_{ij}$ is an angle between two neighboring spins. Terms containing
only $S_i^z$ and/or $S_j^z$ provide the classical energy and the mean-field
fluctuations of spins in a local magnetic field. By analogy to other
frustrated models  \cite{Heinila93,Canals04}, one can show  that a local field
for the TAFM is the same  on every site $H_{\rm loc} = 3J$ irrespective of
$H\leq H_s$. Consequently, the local
mean-field fluctuations are governed by the Hamiltonian
\begin{equation}
\hat{\cal H}_0 = -H_{\rm loc}\sum_i S_i^z \simeq \frac{H_{\rm loc}}{2} \sum_i
\bigl[ (S_i^x)^2 +  (S_i^y)^2 \bigr]   \ ,
\label{H0}
\end{equation}
where  we expanded $S_i^z=\sqrt{1-(S_i^x)^2-(S_i^y)^2}$ and use $\simeq$ to indicate
that a  constant is dropped in the final expression.

In the absence of random disorder ($J_{ij}\equiv J$) the terms
linear in $S_i^x$ sum up to zero for all classical ground states.
Then, the perturbation to Eq.~(\ref{H0}) is
\begin{equation}
\hat{V}_1 =  J \sum_{\langle ij\rangle}
\bigl(  S_i^y S_j^y + S_i^x S_j^x  \cos\theta_{ij}  \bigr)  \ .
\label{V1}
\end{equation}
The leading correction  to the free energy
is given by $\Delta F = -\langle\hat{V}_1^2\rangle/2T$. Using
$\langle (S_i^{x})^2\rangle = \langle (S_i^y)^2\rangle = T/H_{\rm loc}$
derived from Eq.~(\ref{H0}), we obtain
\begin{equation}
\Delta F = - \frac{J^2T}{2H_{\rm loc}^2} \sum_{\langle ij\rangle}
( \cos^{2}\theta_{ij}\!+1)  \simeq -\frac{T}{18}\sum_{\langle ij\rangle}
(\mathbf{S}_{i} \cdot\mathbf{S}_{j}) ^{2} \,,
\label{DF}
\end{equation}
where in the last expression we restore the mean-field (ground-state) spin directions.
Thus, short-wavelength thermal fluctuations produce an effective biquadratic exchange.
Quantum fluctuations also  generate a similar term \cite{Heinila93}.
Because of its negative sign, the biquadratic exchange (\ref{DF}) favors the ``most collinear''
spin configurations among degenerate classical states. For the TAFM
this leads to selection of the coplanar configurations, Figs.~\ref{fig:conf}(a)
and \ref{fig:conf}(c), at low and high fields, respectively, and to appearance
of the 1/3 magnetization plateau with the collinear up-up-down ($uud$)
spin structure, Fig.~\ref{fig:conf}(b).

We now set $T=0$ and consider the effect of quenched disorder, which locally
violates the perfect geometrical frustration. In this case, the linear terms
provide the  main perturbation to the classical energy
\begin{equation}
\hat{V}_2 =  \sum_{\langle ij\rangle} \delta J_{ij} \sin\theta_{ij} (S_j^x - S_i^x)  \ .
\label{V2}
\end{equation}
Minimization of $\hat{\cal H}_0 + \hat{V}_2$ with respect to $S_i^x$
under the assumption that bonds fluctuate independently yields
\begin{eqnarray}
\Delta E = - \frac{\delta J^2}{2H_{\rm loc}} \sum_{i,j} \sin^2 \theta_{ij} \simeq
\frac{\delta J^2}{3J} \sum_{\langle ij\rangle} (\mathbf{S}_i \cdot \mathbf{S}_j)^2 \ ,
\label{DE}
\end{eqnarray}
The energy correction generated by the bond disorder has the same functional form
as (\ref{DF}) but with the opposite sign. Therefore, the configurational
disorder favors the ``least collinear'' states in the ensemble of degenerate classical
ground sates. Selection of orthogonal or ``anticollinear'' ground states was previously known
in the context of the diluted $J_1$--$J_2$ square-lattice antiferromagnet
\cite{Henley89,Weber12} yet the tendency determined by (\ref{DE}) is rather general,
see also Refs.~\cite{Fyodorov91,Larson08} with similar conclusions.

In the case of the Heisenberg TAFM in an external field the  least collinear state
corresponds to the conical spin structure,
Fig.~\ref{fig:conf}(d). Thus, the two types of disorder, thermal and quenched,
compete with each other producing a rich $H$--$T$ phase diagram.
Note that Eqs.~(\ref{DF}) and (\ref{DE}) are only approximate
and constitute the first terms in the $1/z$ expansion, $z$ being the number of the nearest
neighbors \cite{Larson08}. Still, as comparison with the numerics
shows, the effective biquadratic exchange is able to capture the principal qualitative
tendencies for the TAFM.

{\it Ground state selection.}---%
To extend the qualitative analytical result obtained for the weak bond disorder
to the case of random vacancies we performed numerical minimization of
the classical energy (\ref{H}). The minimization is carried out for periodic
$L\times L$ clusters with fixed concentration of nonmagnetic impurities
$n_{\rm imp} = 0.1$--5\% and linear sizes up to $L=150$. One starts with a random
spin configuration and solves iteratively the local minimum condition
${\bf S}_i^{(k)}\parallel {\bf h}_i^{(k)}$, where the local field is
${\bf h}_i^{(k)} = {\bf H} -J \sum_j p_j{\bf S}_j^{(k-1)}$.  Once converged
the procedure is repeated with up to $10^3$ random initial
configurations and the global minimum is selected. Physical quantities are then
averaged over 100 impurity replicas.

\begin{figure}[t]
\includegraphics[width=0.99\columnwidth, keepaspectratio]{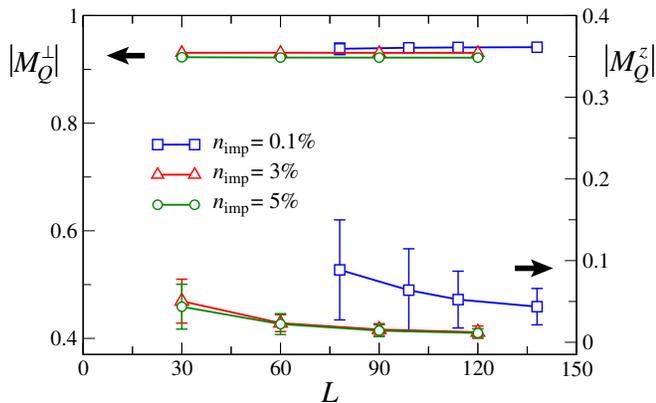}  
\caption{Zero-temperature transverse $|M_{Q}^{\perp}|$ and longitudinal $|M_{Q}^{z}|$
antiferromagnetic order parameters at $H/J=3$ for clusters with different
concentration of vacancies $n_{\rm imp}$ and different linear size $L$.
}
\label{fig:GS}
\end{figure}

Ground-state configurations of the TAFM in magnetic field are characterized by
the antiferromagnetic order parameter:
\begin{equation}
\mathbf{M}_Q = \frac{1}{N} \sum_i  \mathbf{S}_{i}e^{-i\mathbf{Q}{\bf r}_i}
\label{OPs}
\end{equation}
with $N$ being the number of filled sites.
In particular, the conical state is unambiguously distinguished from the coplanar
configurations
by a finite $M^\perp_{Q}=(|M^{x}_{Q}|^2+ |M^{y}_{Q}|^2)^{1/2}$ and $|M^z_{Q}|=0$.

Numerical results for transverse and longitudinal components of the AFM order parameter
at $H/J=3$ are shown in Fig.~\ref{fig:GS}. The conical state remains stable
for all studied impurity concentrations including the smallest one $n_{\rm imp}=0.1$\%.
The lack of appreciable finite-size effects in $M_Q^{\perp}$  indicates
the absence of a spin-glass phase and the development of
the true long-range order in transverse components. A similar behavior is found for
all $0<H<H_s$ albeit with more iteration steps
required for $H\to 0$.  Hence,  the numerical
results for the diluted TAFM fully corroborate the analytical findings
for the bond-disorder model. The vacancy moment effect \cite{Wollny11,Wollny12} averages
to zero due to equal occupation of magnetic sublattices by impurities and the
finite-field behavior is determined by configurational fluctuations which are correctly captured
by the bond-disorder model.

{\it Phase diagram.}---%
We have performed the classical Monte Carlo (MC) simulations of the diluted TAFM
in a wide range of temperatures and magnetic fields using the hybrid algorithm,
which combines the Metropolis step with a few over-relaxation moves, see
\cite{MZH08,Gvozdikova11} for further details.
Physical quantities and associated error bars were estimated from averaging over
$100$ impurity replicas. Phase transition boundaries
were determined by the standard finite-size scaling analysis of the
the fourth-order Binder cumulants for the AFM order parameter
(\ref{OPs}) and the associated spin chirality,
as well as from the behavior of the spin
stiffness  and the specific heat on clusters with linear sizes up to $L=150$.

\begin{figure}[b]
\includegraphics[width=0.9\columnwidth, keepaspectratio]{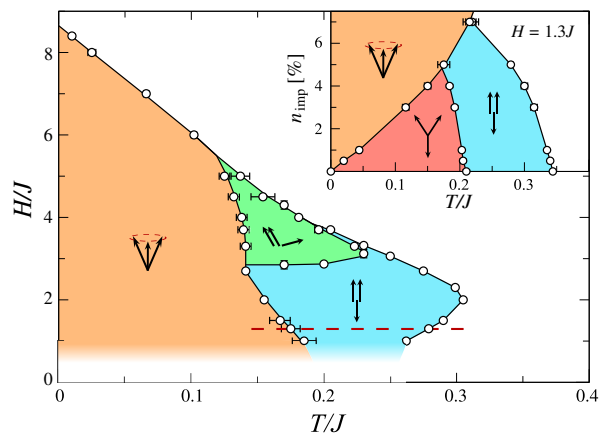}   
\caption{
Classical Monte Carlo phase diagram of the Heisenberg TAFM
with 5\% of nonmagnetic impurities.
Solid lines via data points are guides for the eye.
The inset shows the concentration evolution of ordered
phases for $H/J=1.3$, which is indicated by the dashed line
on the main panel.
}
\label{fig:PhD}
\end{figure}

The magnetic phase diagram of the Heisenberg TAFM with 5\% of vacancies is shown
in Fig.~\ref{fig:PhD}. The main new feature in comparison with the diagram
of the pure model \cite{Kawamura85,Gvozdikova11,Seabra11} is the emergence of
the conical state at low temperatures for all  $H \leq H_s$. At high enough temperatures
the increased thermal fluctuations overcome the quenched disorder and magnetic phases
of the pure TAFM reappear again, though the $Y$-phase remains absent for $n_{\rm imp}=5$\%.
The phase transition boundaries are drawn in Fig.~\ref{fig:PhD} down to $H/J\sim 1$.
In lower magnetic fields,  the finite-size effects become stronger and require
simulations of significantly larger clusters than those studied in our work.
Therefore, we cannot exclude reappearance of the $Y$-phase at very low fields. Instead
we show evolution of the phase boundaries
with the vacancy concentration at fixed $H=1.3J$ in the inset of Fig.~\ref{fig:PhD}.
The $Y$-phase is present in this field for small $n_{\rm imp}$ and disappears
at $n_{\rm imp}\sim 4.5\%$.

The effect of impurities on the critical behavior of the TAFM can
be assessed using the Harris criterion \cite{Harris74}, which states that the
disorder becomes relevant for transitions with $\alpha = 2 - d\nu>0$. In particular,
the Berezinskii-Kostelitz-Thouless (BKT) transition formally has $\nu=\infty$
and remains unaffected by vacancies as was confirmed  numerically in
\cite{Berche03}. The second-order transition into the $uud$ state, which belongs
to the universality class of the 2D three-state Potts model,  has $\alpha=1/3$
and is, therefore, driven by impurities to a new random fixed point, see \cite{Kim96,Picco96}
and references therein. Nevertheless, the  spin correlation exponent $\eta$ stays very close
to the clean value $\eta = 4/15$  \cite{Kim96} and we also found virtually no difference
with the pure case for the critical behavior of the order parameter at this transition
in our Monte Carlo simulations.

In the high-field region $5J \alt H < H_{s}$ the direct transition
between paramagnetic and conical states is accompanied by breaking of
the $\mathbb{Z}_2 \otimes O(2)$ symmetry, where  $\mathbb{Z}_2$ describes
the chirality ordering. The statistical errors
in simulations brought by the impurities are too large
to resolve a presumably tiny splitting of $T_{\rm BKT}$ and $T_{\rm chir}$
as well as an effect of the disorder on the Ising-like chiral transition.
Finally, transitions between coplanar states and the conical phase
are expected to be of the first order on symmetry grounds.
Indeed, there is a signature of the first-order transition
for $2J \alt H < 3J$ from the scaling of the specific heat anomaly.
At higher and lower fields the diluted TAFM shows fingerprints of a continuous
transition between conical and coplanar states, which may also signify
presence of an intermediate phase in a narrow temperature interval.

{\it XY triangular antiferromagnet.}---%
Let us now briefly discuss
the effect of nonmagnetic impurities in the easy-plane TAFM---a model relevant
to a number of real materials. Some of them exhibit the 1/3 magnetization plateau for
fields applied parallel to the easy plane \cite{Kenzelmann07,Susuki13},
which is a clear sign of geometrical magnetic
frustration.
The ordered states of the $XY$ TAFM in the presence of an in-plane magnetic
field were investigated by Lee {\it et al.}\ \cite{Lee84}.
Thermal fluctuations lift the  ground-state degeneracy in favor of the same
sequence of phases in magnetic field as for the Heisenberg model,
see Fig.~\ref{fig:conf}(a)--(c).
Our derivation of an effective biquadratic exchange for the weak bond disorder
remains intact for the $XY$ spins.
Hence, the only difference with the isotropic case is that the conical state,
Fig.~\ref{fig:conf}(d), as well as other non-coplanar configurations are
now forbidden. Therefore, the biquadratic
exchange (\ref{DE}) lifts the degeneracy between
the coplanar structures only. An elementary analysis shows
that the lowest energy state favored by a positive biquadratic exchange
corresponds to the ``anti-$Y$' spin configuration shown in Fig.~\ref{fig:conf}(e). In
stronger magnetic fields
the two canted spins tilt further  towards the field direction
continuously transforming the  anti-$Y$ state into the fan spin structure,
Fig.~\ref{fig:conf}(f).

We have complemented analytical consideration with numerical search
for the lowest-energy magnetic structures using the same technique as
for the isotropic model. Numerical results, which
will be reported elsewhere \cite{SM}, are fully consistent with the presence
of the fan (anti-$Y$) state in the whole range of magnetic fields.
Interestingly, a new high-field state interpreted as a fan
structure was recently observed in the easy-plane spin-1/2 TAFM $\rm Ba_3CoSb_2O_9$
\cite{Susuki13}. Though the full theoretical explanation of the new phase
should include quantum effects, our analysis of the classical model suggests that
nonmagnetic impurities may play a key role in its appearance.

{\it Discussion.}---%
Nonmagnetic impurities modify the behavior of the classical TAFM in an external
magnetic field. The effect of static disorder can be qualitatively described by a positive
biquadratic exchange, which competes with a similar effective interaction of
the opposite sign generated by thermal and quantum fluctuations. At zero temperature
vacancies stabilize the conical state for the Heisenberg TAFM, whereas for
the $XY$ model with an in-plane field they favor the fanlike spin structure. A
similar competition between the quenched disorder and thermal effects must
also be present in other geometrically frustrated antiferromagnets.

Beyond the classical model, quantum fluctuations will compete with the effect of
dilution even at $T=0$. For a given spin value $S$, there is a critical concentration
of vacancies $n^c_{\rm imp}\sim 1/S$ needed to overcome the quantum selection
of `most collinear' states.
Comparing the harmonic spin-wave energies
of the $uud$ and conical states \cite{Chubukov91} with the classical
energy gain of the conical state obtained from the numerical minimization we
find that the 1/3 magnetization plateau of the Heisenberg TAFM
is stable up to $n^c_{\rm imp}\sim 4$\% for $S=5/2$  \cite{SM}.
This estimate for $n^c_{\rm imp}$ becomes even lower
once quantum effects are further suppressed by weak magnetic anisotropy.
Systematic experimental studies of frustrated magnets doped with
nonmagnetic impurities may, therefore,  bring new fascinating physics.
Apart from  the fundamental interest, this can open additional
possibilities in controlling electrical and magnetic polarizations in
triangular multiferroics.

We acknowledge helpful discussions with A. L. Chernyshev and M. Vojta.


\begin{thebibliography}{99}



\bibitem{Vajk02}
O. P. Vajk, P. K. Mang, M. Greven, P. M. Gehring, and J. W. Lynn,
Science {\bf 295}, 1691 (2002).

\bibitem{Bobroff09}
J. Bobroff {\it et al.},
Phys. Rev. Lett. {\bf 103}, 047201 (2009).

\bibitem{Sigrist96}
M. Sigrist and A. Furusaki,
J. Phys. Soc. Jpn. {\bf 65}, 2385 (1996).

\bibitem{Sachdev99}
S. Sachdev, C. Buragohain, and M. Vojta,
Science {\bf 286}, 2479 (1999).

\bibitem{Sachdev03}
S. Sachdev and M. Vojta,
Phys. Rev. B {\bf 68}, 064419 (2003).

\bibitem{Sushkov03}
O. P. Sushkov,
Phys. Rev. B {\bf 68}, 094426 (2003).

\bibitem{Hoglund07}
K. H. H\"oglund, A. W. Sandvik, and S. Sachdev,
Phys. Rev. Lett. {\bf 98}, 087203 (2007).

\bibitem{Eggert07}
S. Eggert, O. F. Sylju\aa sen, F. Anfuso, and M. Anders,
Phys. Rev. Lett. {\bf 99}, 097204 (2007).

\bibitem{Chen11}
C.-C. Chen, R. Applegate, B. Moritz, T. P. Devereaux, and R. R. P. Singh,
New J. Phys. {\bf 13}, 043025 (2011).

\bibitem{Sen11}
A. Sen, K. Damle, and R. Moessner,
Phys. Rev. Lett. {\bf 106}, 127203 (2011).

\bibitem{Wollny11}
A. Wollny, L. Fritz, and M. Vojta,
Phys. Rev. Lett. {\bf 107}, 137204 (2011).

\bibitem{Weber12}
C. Weber and F. Mila,
Phys. Rev. B {\bf 86}, 184432 (2012).

\bibitem{Wollny12}
A. Wollny, E. C. Andrade, and M. Vojta,
Phys. Rev. Lett. {\bf 109}, 177203 (2012).

\bibitem{Brenig13}
W. Brenig and A. L. Chernyshev,
Phys. Rev. Lett. {\bf 110}, 157203 (2013).

\bibitem{Villain79}
J. Villain, Z. Phys. B {\bf 33}, 31 (1979).

\bibitem{Henley89}
C. L. Henley,
Phys. Rev. Lett. {\bf 62}, 2056 (1989).

\bibitem{Savary11}
L. Savary, E. Gull, S. Trebst, J. Alicea, D. Bergman, and L. Balents,
Phys. Rev. B {\bf 84}, 064438 (2011).

\bibitem{Lee84}
D. H. Lee, J. D. Joannopoulos, J. W. Negele, and D. P. Landau,
Phys. Rev. Lett. {\bf 52}, 433 (1984).

\bibitem{Kawamura85}
H. Kawamura and S. Miyashita,
J. Phys. Soc. Jpn. {\bf 54}, 4530 (1985).

\bibitem{Chubukov91}
A.~V.~Chubukov and D.~I.~Golosov,
J.~Phys.\ Condens.\ Matter {\bf 3}, 69 (1991).

\bibitem{Balents10}
L. Balents, Nature \textbf{464}, 199 (2010).

\bibitem{Struck11}
J. Struck  {\it et al.}, Science \textbf{333}, 996 (2011).

\bibitem{Katsufuji01}
T. Katsufuji, S. Mori, M. Masaki, Y. Moritomo, N. Yamamoto, and H. Takagi,
Phys. Rev. B {\bf 64}, 104419 (2001).

\bibitem{Kenzelmann07}
M. Kenzelmann {\it et al.},
Phys. Rev. Lett. {\bf 98}, 267205 (2007).

\bibitem{Long89}
M. W. Long, J. Phys.: Condens. Matter {\bf 1}, 2857 (1989).

\bibitem{Heinila93}
M. T. Heinil\"a and A. S. Oja,
Phys. Rev. B {\bf 48}, 7227 (1993).

\bibitem{Canals04}
B. Canals and M. E. Zhitomirsky,
J. Phys.: Condens. Matter {\bf 16}, S759 (2004).

\bibitem{Fyodorov91}
Y. V. Fyodorov and E. F. Shender,
J. Phys.: Condens. Matter {\bf 3}, 9123 (1991).

\bibitem{Larson08}
B. E. Larson and C. L. Henley, arXiv:0811.0955 (unpublished).

\bibitem{MZH08}
M. E. Zhitomirsky,
Phys. Rev. B {\bf 78}, 094423 (2008).

\bibitem{Gvozdikova11}
M. V. Gvozdikova, P.-E. Melchy, and M. E. Zhitomirsky,
J. Phys.: Condens. Mat. {\bf 23}, 164209 (2011).


\bibitem{Seabra11}
L. Seabra, T. Momoi, P. Sindzingre, and N. Shannon,
Phys. Rev. B {\bf 84}, 214418 (2011).

\bibitem{Harris74}
A. B. Harris,
J. Phys. C {\bf 7}, 1671 (1974).

\bibitem{Berche03}
B. Berche, A. I. Farinas-Sanchez, Yu. Holovatch, and R. Paredes V.,
Eur. Phys. J. B {\bf 36}, 91 (2003).

\bibitem{Kim96}
J.-K. Kim, Phys. Rev. B {\bf 53}, 3388 (1996).

\bibitem{Picco96}
M. Picco, Phys. Rev. B {\bf 54}, 14930 (1996).

\bibitem{Susuki13}
T. Susuki {\it et al.},
Phys. Rev. Lett. {\bf 110}, 267201 (2013).


\bibitem{SM}
V. S. Maryasin and M. E. Zhitomirsky,  to be published.

\end{thebibliography}
\end{document}